\documentclass[aps,prl,showpacs,notitlepage,floatfix,twocolumn,superscriptaddress]{revtex4-2}
\usepackage{amssymb}
\usepackage{amsmath}
\usepackage{amsfonts}
\usepackage{bbm}
\usepackage{graphicx}
\usepackage{braket}
\usepackage{mathtools}
\usepackage{dsfont}
\usepackage{physics}
\usepackage{bbold}
\usepackage[dvipsnames]{xcolor}
\usepackage[titletoc]{appendix}

\newcommand\myshade{85}
\colorlet{myurlcolor}{MidnightBlue}

\usepackage[normalem]{ulem}

\usepackage[colorlinks,linkcolor=myurlcolor!\myshade!black,anchorcolor=myurlcolor!\myshade!black,citecolor=myurlcolor!\myshade!black,urlcolor=MidnightBlue]{hyperref}

\def\I{\mathcal{M}}

\def\bea{\begin{equation}\begin{aligned}}
\def\eea{\end{aligned}\end{equation}}

\begin{document}
\setlength{\unitlength}{1mm}

\title{Rainbow Scars: From Area to Volume Law}
\author{Christopher M. Langlett}
\affiliation{Department of Physics \& Astronomy, Texas A\&M University, College Station, Texas 77843, USA}

\author{Zhi-Cheng Yang}
\affiliation{Joint Center for Quantum Information and Computer Science, NIST/University of Maryland, College Park, Maryland 20742, USA}
\affiliation{Joint Quantum Institute, NIST/University of Maryland, College Park, Maryland 20742, USA}

\author{Julia Wildeboer}
\affiliation{Department of Physics \& Astronomy, Iowa State University, Ames, Iowa 50011, USA}

\author{Alexey V. Gorshkov}
\affiliation{Joint Center for Quantum Information and Computer Science, NIST/University of Maryland, College Park, Maryland 20742, USA}
\affiliation{Joint Quantum Institute, NIST/University of Maryland, College Park, Maryland 20742, USA}

\author{Thomas Iadecola}
\email{iadecola@iastate.edu}
\affiliation{Department of Physics \& Astronomy, Iowa State University, Ames, Iowa 50011, USA}

\author{Shenglong Xu}
\email{slxu@tamu.edu}
\affiliation{Department of Physics \& Astronomy, Texas A\&M University, College Station, Texas 77843, USA}

\begin{abstract}
Quantum many-body scars~(QMBS) constitute a new quantum dynamical regime in which rare ``scarred" eigenstates mediate weak ergodicity breaking.
One open question is to understand the most general setting in which these states arise.
In this work, we develop a generic construction that embeds a new class of QMBS, rainbow scars, into the spectrum of an \textit{arbitrary} Hamiltonian.
Unlike other examples of QMBS, rainbow scars display extensive bipartite entanglement entropy while retaining a simple entanglement structure.
Specifically, the entanglement scaling is volume-law for a random bipartition, while scaling for a fine-tuned bipartition is sub-extensive.
When internal symmetries are present, the construction leads to multiple, and even towers of rainbow scars revealed through distinctive non-thermal dynamics.
Remarkably, certain symmetries can lead rainbow scars to arise in translation-invariant models.
To this end, we provide an experimental road map for realizing rainbow scar states in a Rydberg-atom quantum simulator, leading to coherent oscillations distinct from the strictly sub-volume-law QMBS previously realized in the same system.
\end{abstract}
\maketitle

Statistical mechanics relies on relaxation towards the maximally entropic state in thermal equilibrium.
This process, however, is at odds with the fact that the entropy of a many-body system prepared in a pure state must remain identically zero under unitary dynamics.
The emergence of statistical mechanics in such systems, known as quantum thermalization, proceeds by the relaxation of local sub-regions to a thermal state via the exchange of quantum correlations with the remainder of the system.
This mechanism, whereby a pure state becomes locally indistinguishable from a thermal state, follows from the eigenstate thermalization hypothesis~(ETH)~\cite{ETH1, ETH2, ETH3, ETH4}.
The ETH postulates a correspondence between the local reduced density matrix of a finite-energy-density eigenstate and the Gibbs ensemble.

Many lines of inquiry involve constructing systems where thermalization is avoided.
For example, quantum integrable systems~\cite{qIntegrable1,qIntegrable2} fail to thermalize due to extensively many conservation laws; however, these systems are unstable to perturbations.
A more robust violation of the ETH arises in disordered interacting systems, which may induce many-body localization, resulting in an extensive number of conservation laws~\cite{mbl1,mbl2,mbl3,mbl4}. 

Experiments utilizing cold atoms~\cite{expt1, expt2,expt4, expt8,expt3,expt5,expt11}, ion traps~\cite{expt6, expt9}, and superconducting circuits~\cite{expt7, expt10} have demonstrated unprecedented control over the dynamics of many-body systems.
Recently, experiments in Rydberg-atom arrays simulating quantum Ising models in varying dimensions~\cite{scarexpt1,scarexpt2} observed sustained coherent oscillations of local observables for special initial states, such as the N\'eel state.
This observation was later traced to the existence of rare, weakly entangled eigenstates in an otherwise thermal system~\cite{scartheo2,scartheo2.1}.
This phenomenology was dubbed ``quantum many-body scars''  (QMBS)~\cite{serbyn2021_review}, an earlier example of which was found in the Affleck-Kennedy-Lieb-Tasaki spin chain in Refs.~\cite{aklt1,aklt2}.
QMBS have been studied in a wide range of systems, including the ``PXP model'' simulated by the Rydberg experiment~\cite{scartheo3,scartheo4,scartheo8,scartheo9,Iadecola2019_magnon}, the spin-1 XY model~\cite{scartheo5,scartheo16}, Fermi-Hubbard models~\cite{scartheo17,motrunich}, Floquet models~\cite{scartheo10,scartheo12,scartheo13,Mukherjee2020_floquet,automaton}, and other systems~\cite{thintorus,scartheo7,scartheo14,scartheo11,scartheo18,scartheo19,Yoshi2020_flat,McClarty2020_frustration,Banerjee2021_gauge,hfrag1,Zhao21}.
Group-theoretic techniques~\cite{scarsym1,scarsym2,scarsym3,scartheo17,scarsym4,scarsym5}, matrix product state methods~\cite{scar-mps}, and projector embeddings~\cite{embed1,embed2} have been employed to systematically generate sub-volume-law QMBS in the many-body spectrum.
It remains an open question to construct QMBS with a specific entanglement structure in the spectrum of a generic system.

In this work, we develop a general construction for a new class of QMBS, rainbow scars~\cite{rainbow1, rainbow2, rainbow3}, in the spectrum of an \textit{arbitrary} Hamiltonian governing a replicated system.
Rainbow scars differ from previous examples of QMBS in that their entanglement scaling strongly depends on the chosen bipartition.
Specifically, the entanglement is volume-law for a random cut, but sub-volume-law for a fine-tuned cut.
In the presence of symmetries, multiple and even towers of rainbow scar states emerge, and may exhibit a rich group theoretic structure.
This opens the possibility to probe the scar states with quantum quenches.
Furthermore, certain symmetries can even yield rainbow scars in simple translation-invariant models.
We propose a realization of rainbow scars in a system of interacting Rydberg atoms, where these states lead to coherent oscillatory dynamics whose origin is fundamentally distinct from the previously studied sub-volume law QMBS.

\textit{General Construction.}---Imagine two related copies of a quantum many-body system with the Hamiltonian:
\bea\label{eq:general_construction} 
H = H_{1}\otimes \mathbbm{1}+\mathbbm{1}\otimes H_{2}+\lambda_{\text{c}} V_{\text{c}}.
\eea
Each subsystem $H_1$ and $H_2$ consists of $N$ sites with a $d$-dimensional local Hilbert space, spanned by the local computational basis $\ket{s_i}$ at site $i$.
The state $\ket{S}=\prod_i \ket{s_{i}}$ defines the global computational basis spanning a Hilbert space of dimension $d^{2N}$.
Moreover, in 1D~\footnote{The construction Eq.~\eqref{eq:general_construction} is valid for arbitrary dimensions, where the mirror-symmetry operator $\mathcal{M}$ is the map $\mathcal{M}: i\rightarrow \tilde{i}$.
For concreteness we restrict ourselves to one-dimensional systems.}, the ``copied" Hamiltonian, $H_2$, satisfies $H_{2}= -\mathcal{M}H^{*}_{1}\mathcal{M}$, with the mirror-symmetry operator $\mathcal{M}$ mapping $i \rightarrow \tilde{i}\equiv 2N-i+1$.
Complex conjugation is defined with respect to the computational basis $\ket{S}$. 
The two systems interact through $V_{\text{c}}$, which generically thermalizes the combined system, akin to two boxes of gas equilibrating through a thin connecting wire.
Provided the condition $H_2=-\mathcal{M} H_1^* \mathcal{M}$ is met, the construction is independent of the microscopic details of $H_{1(2)}$.
This strict condition on $H_2$ is relaxed in the presence of certain symmetries, as discussed below.
\begin{figure}
    \includegraphics[width=\columnwidth]{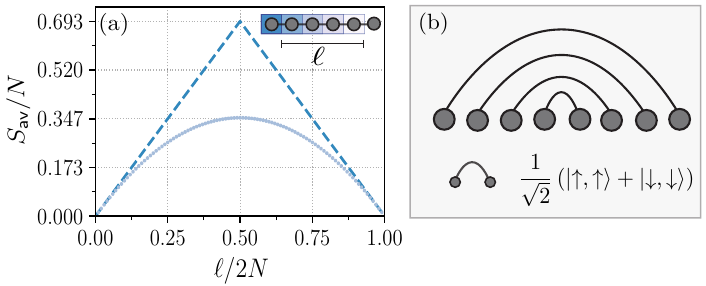}
    \caption{\textit{Entanglement Scaling of Random Bipartition.}
    (a) Average entanglement for each bipartition $\ell \in [0, 2N]$, the dotted line indicates maximal entanglement, here $2N=200$.
    Inset: Depiction of bipartitions.
    (b) Rainbow state for $d=2$ with each bond a Bell state.
    }
    \label{fig:panel1}
\end{figure}

\begin{figure}
    \includegraphics[width=\columnwidth]{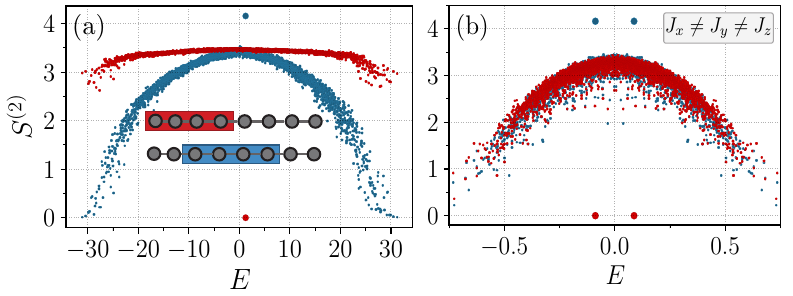}
    \caption{\textit{Second-order R\'enyi Entropy.}
    (a) Second-order R\'enyi entropy for a random Hamiltonian drawn from the GUE with a Heisenberg coupling, $\lambda_{\text{c}}=5.0$.
    inset: Chosen entanglement cuts: standard bipartition~(blue) and fine-tuned bipartition~(red).
    (b) Translation-invariant model with $J_{x}=0.20, J_{y}=0.15, J_{z}= 0.25$ using open boundary conditions.}
    \label{fig:panel1_a}
\end{figure}

We proceed by illustrating how a class of non-thermal states emerges from a large set of degenerate states through a carefully chosen coupling.
Using the spectral decomposition to express $H_{1}=\sum_{n=1}^{d^{N}}E_{n}\ket{\psi_n}\bra{\psi_n}$, where $H_1\ket{\psi_n}=E_{n}\ket{\psi_n}$.
Similarly, express $H_2=-\sum_{n=1}^{d^{N}}E_n \ket{\I \psi_n^*}\bra{\I \psi_n^*}$, where $\ket{\I\psi_n^*}\equiv(\I\ket{\psi_n})^*$.
At $\lambda_\text{c}=0$, the eigenstates of the total Hamiltonian $H$, with eigenvalues $E_n-E_m$, are
$
\{\ket{\Psi_{nm}}=\ket{\psi_n}\otimes \ket{\I \psi_m^*}:\ \forall \, n,m = 1, \ldots, d^{N}\}
$, which have no entanglement between the two halves.
Consequently, $H$ has a $d^{N}$-fold degenerate subspace spanned by $\ket{\Psi_{nn}}$.
Within this degenerate subspace, there exists a special eigenstate independent of the details of $H_1$:
\bea
\label{eq:rainbow_state} 
\ket{I} & = \frac{1}{d^{N/2}}\sum_{n=1}^{d^N}\ket{\Psi_{nn}}=\frac{1}{d^{N/2}} \bigotimes_{i=1}^{N} \sum_{s=0}^{d-1}\ket{s_i}\ket{s_{\tilde{i}}},
\eea
where the second equality follows from inserting a resolution of the identity.
This state is precisely the ``rainbow state''~\cite{rainbow1, rainbow2, rainbow3}, named for its characteristic pattern of entanglement, in which every site $i$ is maximally entangled with its mirror partner $\tilde{i}$~[see Fig.~\ref{fig:panel1}(b) middle inset].
The rainbow state is also known as the infinite-temperature thermofield double state; it is of interest in the high-energy community~\cite{cottrell2019build,hartman2013time,PhysRevLett.115.211601,blackhole1, blackhole2} for its connections to black-hole physics, and in the quantum information community where it is used as an entanglement resource~\cite{schuster2021many,nezami2021quantum,brown2019quantum}. 
The entanglement entropy for the standard bipartition~[see Fig.~\ref{fig:panel1_a}(a) top inset] scales linearly with system size, $S=N\log d$, while retaining a simple structure.
More generally, for a random bipartition defining a sub-region $A$ of size $\ell$, the entanglement scales extensively on average when $\ell\propto N$: $S_{\text{av}}=(2N-\ell)\ell \log(d)/(2N-1)$~[Fig.~\ref{fig:panel1}(a)]~[see Supplementary Material~(SM)~\cite{smaterial}].
The rainbow state is denoted as the state $\ket{I}$ corresponding to the identity operator under the state-channel duality~\cite{channel1, channel2}.
For $\lambda_\text{c}\neq 0$, the rainbow state is selected as an eigenstate of the local Hamiltonian $H$ from the degenerate subspace provided $\ket{I}$ is an eigenstate of $V_\text{c}$.
Specifically, for $d=2$, $\ket{I}$ is a product of long-range Bell states, $\ket{I}=\bigotimes_{i\leq N}\left(\ket{\uparrow,\uparrow}+\ket{\downarrow,\downarrow}\right)_{i, \tilde{i}}$.
If the subsystems are coupled through, e.g., a Heisenberg interaction, $V_{\text{c}}=\Vec{S}_{N}\cdot \Vec{S}_{N+1}$, then $\ket{I}$ is an eigenstate of the combined system with energy $E_{I}=\lambda_{\text{c}}/4$.

To emphasize the generality of the construction, consider a system of $2N$ qubits for which $H_1$~(which fixes $H_2$) is randomly drawn from the Gaussian unitary ensemble~(GUE), with a local Heisenberg coupling acting on the central qubits.
Fig.~\ref{fig:panel1_a}(a) shows the second-order R\'enyi entropy, $S^{(2)}\equiv -\log\text{tr}(\rho_{A}^{2})$, for each eigenstate of $H$, where $\rho_{A}$ is the reduced density matrix of sub-region $A$ for two different entanglement cuts.
Blue points denote the standard bipartition, while the red points denote the fine-tuned bipartition~[see Fig.~\ref{fig:panel1_a}(a) inset].
The appearance of a ``thermalization band''~\cite{thermband1, thermband2, thermband3, thermband4, thermband5} in both cases indicates that the coupling brings the combined system to equilibrium, as expected for a random chaotic model.
Additional evidence is obtained through the average level spacing parameter~\cite{levelspac2,mbl3,levelspac3,levelspac4}, $\langle r \rangle \sim 0.594$, which falls near the GUE random matrix result, $0.60$~\cite{levelspac1}. 
For the standard bipartition, the rainbow state is found as a non-degenerate eigenstate \textit{above} the band with maximal entanglement, markedly distinct from previous examples of QMBS.
By contrast, for the fine-tuned bipartition, the rainbow state is a product state, thus violating expectations from ETH.
A priori, a random chaotic model is not expected to host QMBS; nevertheless, the local Heisenberg coupling between the two copies is responsible for selecting $\ket{I}$ from the degenerate subspace and elevating it to a scar.

\textit{Symmetries.}---
First, we discuss how an appropriate symmetry relaxes the condition on $H_2$.
Consider a system with a spectral-reflection symmetry~\cite{specReflec} implemented by an operator $\mathcal{O}$ satisfying $\{\mathcal{O}, H_1\}=0$.
We can then define $H_2=+\I H_1^* \I$ and the state $\ket{\mathcal{O}}=\left(\mathcal{O}\otimes \mathbb{1} \right) \ket{I}$ as an eigenstate of $H_1\otimes \mathbb{1}+\mathbb{1}\otimes H_2$.
This symmetry can even be used to realize the construction in fully translation-invariant models.
For instance, consider the Hamiltonian
\bea\label{eq:trsl_inv}
H = \sum_{i=1}^{2N-1}J_{x}S_{i}^{x}S_{i+1}^{x}+ J_{y}S_{i}^{y}S_{i+1}^{y}+\sum_{i=1}^{2N-2}J_{z}S_{i}^{z}S_{i+1}^{z}S_{i+2}^{z},
\eea
where $S^{\alpha}_i$ are the standard spin-$\frac{1}{2}$ operators on site $i$.
The Hamiltonian above reduces to the form of Eq.~\eqref{eq:general_construction} through a unitary transformation with the operator $\mathcal{O}=\prod_{i=1}^{N} \sigma^x_{N+i}\sigma^y_{N+i+1}$ which flips the sign of the Hamiltonian on the last $N$ sites.
Here the coupling becomes $\lambda_\text{c} V_\text{c} = J_{x}S_{N}^{x}S_{N+1}^{x}+ J_{y}S_{N}^{y}S_{N+1}^{y}+J_{z}(S_{N-1}^{z}S_{N}^{z}S_{N+1}^{z}-S_{N}^{z}S_{N+1}^{z}S_{N+2}^{z})$, for which the rainbow state $\ket{I}$ is an eigenstate.
As discussed above, the state $\mathcal{O}\ket{I}$ then becomes an eigenstate of Eq.~\eqref{eq:trsl_inv}.
Fig.~\ref{fig:panel1_a}(b) shows $S^{(2)}$ for each eigenstate of Eq.~\eqref{eq:trsl_inv} revealing two rainbow scars; the mechanism for multiple scars is elaborated below. 

Symmetries enrich the construction to yield multiple rainbow scar states, which is why two rainbow scars appear in the previous example.
Let $\mathcal{O^\alpha}$ be symmetry generators satisfying $[H_{1}, \mathcal{O^\alpha}] = 0$.
Then the state $\ket{\mathcal{O^\alpha}}=\left(\mathcal{O^\alpha}\otimes \mathbb{1} \right) \ket{I}$ also belongs to the $d^{N}$-fold degenerate subspace at $\lambda_{\rm c}=0$ and is independent of the details of $H_1$.
Provided the $\ket{\mathcal O^\alpha}$ are eigenstates of $V_{\rm c}$, they will emerge as scars in the spectrum. 
For example, consider the case where $H_{1}$ has a $\mathbb{Z}_{2}$ symmetry generated by $\mathcal{O}^x=\prod_{i\leq N} \sigma^{x}_{i}$, where $\sigma^{x}$ is a Pauli operator.
The result is an additional rainbow state, $\ket{X}=\bigotimes_{i\leq N}\left(\ket{\downarrow,\uparrow}+\ket{\uparrow,\downarrow}\right)_{i, \tilde{i}}$.
If $[H_{1}, \mathcal{O}^{\alpha}] = 0$ for each $\mathcal{O}^{\alpha} = \prod_{i\leq N} \sigma^{\alpha}_{i}$ ($\alpha=\{x,y,z\}$), then a set of orthogonal rainbow scars, $\{ \ket{I},\ket{X},\ket{Y},\ket{Z}\}$ arises in the spectrum.
Moreover, an extensive number of rainbow scars emerge if $H$ possesses a global symmetry or kinetic constraints leading to disconnected sub-sectors.

We examine the consequence of symmetries by studying two coupled XYZ chains of $N$ spins:
\bea\label{eq:heisenberg_xyz}
H_1 & = \sum_{i=1}^{N-1}J_{x}S_{i}^{x}S_{i+1}^{x}+ J_{y}S_{i}^{y}S_{i+1}^{y}+J_{z}S_{i}^{z}S_{i+1}^{z}+\tilde{J}S_{i}^{z}S_{i+2}^{z}
\eea
The next-nearest neighbor interaction $\tilde{J}$ is included to prevent integrability.
$H_2$ is set to $-\I H_1^* \I$, and the chains are coupled by $V_{\text{c}}=\vec{S}_N\cdot \vec{S}_{N+1}$.

If $H_1$ commutes with $\mathcal{O}^\alpha$ for $\alpha=\{x,y,z\}$, then four orthogonal rainbow scar states, $\{\ket{I},\ket{X},\ket{Z},\ket{Y}\}$, emerge as eigenstates of $H$.
The first three states correspond to the triplet states of $V_{\rm c}$ and are degenerate with energy $\lambda_{\text{c}}/4$, while the final state is the singlet state of $V_{\rm c}$ at energy $-3\lambda_{\text{c}}/4$.
\begin{figure}
    \includegraphics[width=\columnwidth]{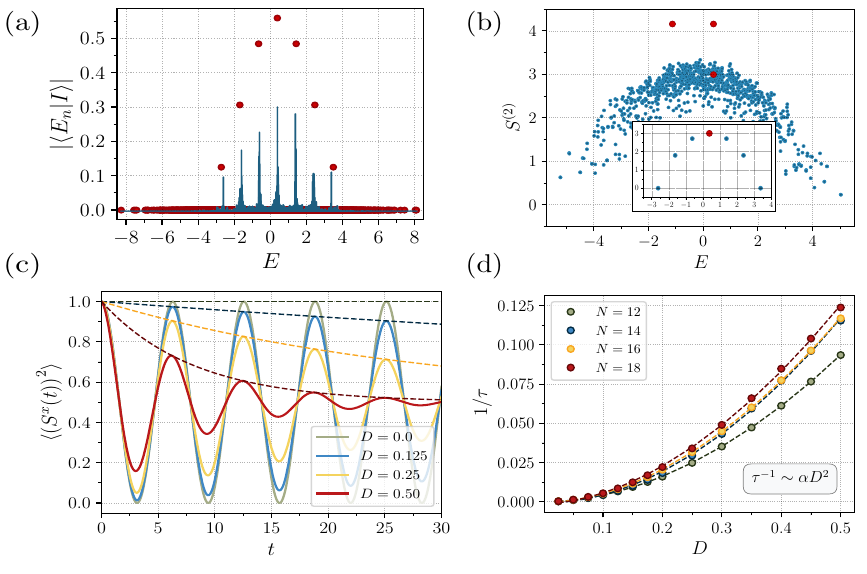}
    \caption{\textit{U(1) tower of rainbow scars.}
    (a) Overlap between the rainbow state $\ket{I}$ and each energy eigenstate of Eq.~\eqref{eq:heisenberg_xyz}, both with~($J_x=J_y=1.0$, red) and without~($J_x=1.0$, $J_y=1.25$, blue) U(1) symmetry.
    (b) Second-order R\'enyi entropy using the standard bipartition within the $S^{z}=0$ sector in the U(1)-symmetric case~($J_{x}=J_{y}=1.5$).
    Inset: The tower highlights the doubly degenerate projected rainbow states $\ket{I}$ and $\ket{Z}$ in each allowed magnetization sector~(red dot indicates $S^{z}=0$).
    (c) Krylov time evolution of $\langle (S^{x}(t))^{2}\rangle/N$ in a system of $2N=18$ spins prepared in $\ket{I}$, with time step $dt=0.1$.
    The dotted lines are fits capturing the amplitude decay.
    (d) Inverse lifetime of $\langle (S^{x}(t))^{2}\rangle/N$ with increasing perturbation strength.
    The remaining parameters used in (a), (b), (c) and (d) are $J_{z}=2.0, \mu=0.5, \Tilde{J}=0.5,\lambda_{\text{c}}=1.5$.
    }
    \label{fig:panel2}
\end{figure}

When $J_{x}=J_{y}$, the total magnetization $S^{z}=\sum^{2N}_{i=1} S^z_i$ of the combined system is conserved.
In this case, the four scars states discussed above are still present, and their projections into each magnetization sector~(if nonzero) are eigenstates.
For instance, the states $\ket{X}$ and $\ket{Y}$ lie within the $S^{z}=0$ sector.
$\ket{I}$ and $\ket{Z}$, instead, have finite projections onto all magnetization sectors with $\sum_{i=1}^N S_i^z = \sum_{i=N+1}^{2N} S_i^z$; these projections coincide up to a global phase, leading to $N+1$ degenerate eigenstates.
Adding $\mu \, S^z$ breaks this degeneracy, resulting in an equally spaced tower of scar states.
This tower of states is created by applying $\hat{J}^{+}=\sum_{i=1}^{N}S_{i}^{+}S_{\tilde{i}}^{+}$ to the fully polarized state $\ket{\Omega}=\bigotimes_{i}\ket{\downarrow}$.
Together with $\hat{J}^{z}=\frac{1}{2}\sum_{i=1}^{2N}S_{i}^{z}$, one can readily verify that the operators $\hat{J}^{\pm}$ and $\hat{J}^{z}$ obey SU(2) commutation relations, so that the tower forms a spin-$N/2$ representation of SU(2).
In Fig.~\ref{fig:panel2}(b), we plot $S^{(2)}$ for each eigenstate with $S^{z}=0$, with the non-thermal states spanning the tower in the inset.
The states $\{\ket{X}, \ket{Y}\}$ in the $S^{z}=0$ sector are non-zero because they are exact eigenstates of the magnetic field term \footnote{We note that if $J_{x}=J_{y}=J_{z}$ the Hamiltonian acquires an SU(2) symmetry leading to a larger rainbow scar tower.
In this circumstance each state is connected with different raising operators satisfying different commutation relations.}. 
In~\cite{smaterial}, we demonstrate that the tower has volume-law entanglement scaling for the standard bipartition and logarithmic scaling for the fine-tuned cut.

Performing a quantum quench from an initial state with finite weight on each eigenstate of the tower leads to perfect coherent dynamics~\cite{scarsym1,scarsym3,scarsym4, scarsym5}.
In particular, preparing Eq.~\eqref{eq:heisenberg_xyz} in either $\ket{I}$ or $\ket{Z}$ results in perfect oscillations, quantified through the non-local correlator, $\langle (S^{x}(t))^{2}\rangle/N$ for $2N=20$ spins, where $S^{x}=\sum S^{x}_i$.
These oscillations are found to be remarkably robust to perturbations.
We perturb Eq.~\eqref{eq:heisenberg_xyz} by setting $J_y-J_x=D$~\cite{scarpert1}; at $D=0$, the U(1) symmetry is exact and the correlator has the analytical form $\langle (S^{x}(t))^{2}\rangle/N = \langle S^{x}(0)^{2}\rangle \cos^{2}(\mu t)/N$.
For $D\neq 0$, the U(1) symmetry is explicitly broken;
yet, the oscillations remain strong for deviations up to $D\sim 0.50$, upon which thermalization sets in~[see Fig.~\ref{fig:panel2}(c)].
We find that the inverse lifetime $1/ \tau \sim \alpha D^{2}$, where $\alpha \approx 0.40$, as expected from Fermi's golden rule~[see Fig.~\ref{fig:panel2}(d)].
Perturbations like $D$ that preserve the structure of Eq.~\eqref{eq:general_construction} yield a more robust dynamical signature than perturbations that break not only U(1) but also the form of Eq.~\eqref{eq:general_construction}~\cite{smaterial}.

\begin{figure}
    \includegraphics[width=\columnwidth]{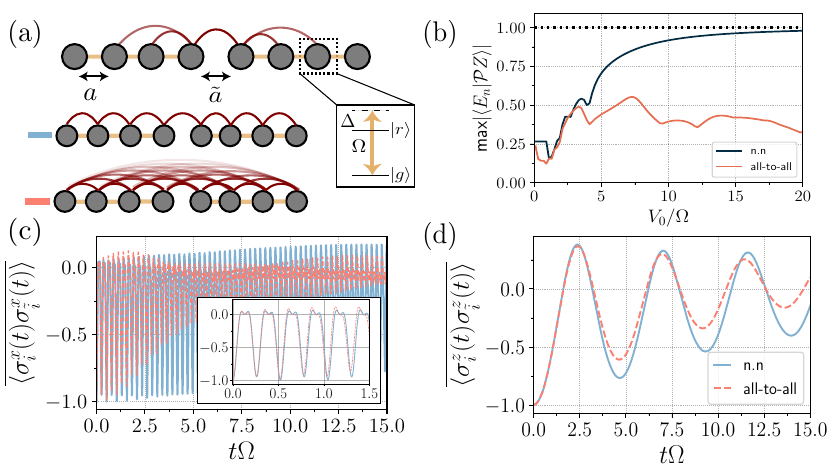}
    \caption{\textit{Dynamical signature in a chain of interacting Rydberg atoms.}
    (a) Depiction of a non-uniformly spaced Rydberg chain.
    (b) Maximum overlap of $\ket{Z}$ projected into the sub-sector absent of neighboring Rydberg states for different interaction strengths.
    Nearest-neighbor~(all-to-all) interactions is denoted by blue(red).
    (c) Dynamics of the average expectation value between inversion pairs, $\overline{\langle \sigma^x_i(t) \sigma^x_{\tilde i}(t)\rangle}$ prepared in $\ket{Z}$.
    Inset: Short time dynamics for $t\Omega\sim 1.5$.
    (d) N\'eel state dynamics for the correlator, $\overline{\langle \sigma^z_i(t) \sigma^z_{\tilde i}(t)\rangle}$.
    Parameters used in (b), (c) and (d): $\Omega / 2\pi = 2$MHz, $V_{0}=12\Omega$, $\Delta_{\text{opt}} = V_{0}/2 \tilde{a}^{6}$ with $\tilde{a}\sim 1.51$ and $2N=16$.}
    \label{fig:panel3}
\end{figure}

\textit{Experimental Realization.}---
As a physically motivating example, we consider a chain of interacting Rydberg atoms with a non-uniform spacing~[see Fig.~\ref{fig:panel3}(a)] governed by the Hamiltonian 
\bea
H =& \frac{\Omega}{2}\sum_{i=1}^{2N}\sigma^{x}_{i}+\sum_{i<j}V_{i,j}n_{i}n_{j}
-\sum_{i=1}^{2N} \Delta_i n_{i}\,.
\label{eq:ham_rydberg}
\eea
Here, we set the interatomic spacing $a=1$ except between sites $N$ and $N+1$, where the spacing is $\tilde{a}$.
The operator $\sigma^{x}_{i}$ connects the internal ground state $\ket{g}_{i}$ to the Rydberg state $\ket{r}_{i}$ of the $i$-th atom, with parameters $\Omega$~(Rabi frequency) and $\Delta_i$~(detuning) characterizing the drive laser.
Rydberg states interact through $V_{i,j}=V_{0}/r_{i,j}^{6}$, with operators $n_{i}=\left(1+\sigma^{z}_{i}\right)/2$.
In the limit $V_{i,i+1}\gg\Omega\gg V_{i,i+2}$, we take $V_{N,N+1}=V_{0}/ \tilde{a}^{6}$ to be comparable to $\Omega$; equivalently, we take $\tilde{a} > 1.0$.
In addition, we take $\Delta_i=0$ except for the two central sites, where $\Delta_{N}=\Delta_{N+1}=\Delta_{\text{opt}}=V_{N,N+1}/2$.
The coupling then becomes $V_0\sigma_{N}^z\sigma_{N+1}^z/4\tilde{a}^{6}$.

In the limit $V_{i,i+1}\gg\Omega\gg V_{i,i+2}$, a pair of U(1) conservation laws emerge, with generators $n^{rr}_{1(2)} = \sum^N_{i(\tilde i)=1}n_{i(\tilde i)}n_{i(\tilde i)+1}$ that count the number of nearest-neighbor pairs of Rydberg excitations in each half of the chain.
The projection of $H$ onto a sector with fixed $n^{rr}_{1,2}$ reads
\bea
\label{eq:rydberg_eff}
H&=H_1+H_2+\frac{V_0}{4\tilde{a}^6}\sigma^z_N \sigma^z_{N+1}+V_0 \left(n^{rr}_1 + n^{rr}_2\right), \\
\eea
with $H_{1(2)}=\mathcal{P}_{1(2)} \left ( \frac{\Omega}{2}\sum_{i=1}^N \sigma^x_i \right) \mathcal{P}_{1(2)}$ where $\mathcal{P}_{1(2)}$ projects the left~(right) half of the chain into a sector with fixed $n^{rr}_{1(2)}$.
The Hamiltonians $H_{1(2)}$ individually have a spectral-reflection symmetry, since $\{\mathcal{O}^{z}, H_{1(2)} \}=0$~\footnote{Here the projection operators, $\mathcal{P}_{1(2)}$ commute with the spectral-symmetry generators, i.e., $[\mathcal{P}_{1(2)}, \mathcal{O}^{z}]=0$.}
When $\mathcal{P}_1=\I \mathcal{P}_2 \I$~(note
$\mathcal{P}_{1}^{*}=\mathcal{P}_{1}$), then $n^{rr}_{1}=n^{rr}_{2}$ and $H_2=+\I H_1^* \I$.
Together with the spectral-reflection symmetry, this implies that the rainbow state $\left(\mathcal{P}_1 \otimes \mathcal{P}_2\right)\ket{Z}$ is an eigenstate of $H_1+H_2$.
This state is also an eigenstate of the coupling, and therefore of the overall $H$ in Eq.~\eqref{eq:rydberg_eff}.
Such a rainbow state exists for each sub-sector satisfying $\mathcal{P}_1=\I \mathcal{P}_2 \I$,
leading to an equally-spaced tower of scar states with energies $V_0/4\tilde{a}^6+2V_0\, n_1^{rr}$.
We emphasize this tower is \textit{distinct} from the strictly sub-volume-law scars of the PXP model, which reside in the sector with $n^{rr}_{1}=n^{rr}_{2}=0$~\cite{serbyn2021_review,scartheo2}.
This tower of states becomes exact in the limit $V_{i,i+1}\gg\Omega\gg V_{i,i+2}$; remarkably, it is also robust away from this limit.

In Fig.~\ref{fig:panel3}(b), we determine the maximum overlap between each eigenstate and the projection of $\ket{Z}$ into the $n^{rr}_{1}=n^{rr}_{2}=0$ sector.
For strictly nearest-neighbor interactions~(blue), the maximum overlap asymptotes to unity as $V_{0}\to\infty$.
However, this is not the case when the full van der Waals interaction is accounted for~(red); here, the overlap grows slowly, never exceeding $\sim 0.5$.
This is a result of the next-nearest-neighbor interactions breaking the spectral-reflection symmetry of $H_{1,2}$ in Eq.~\eqref{eq:rydberg_eff}. 

Fig.~\ref{fig:panel3}(c) shows the quench dynamics of the $\ket{Z}$ rainbow state under the Hamiltonian~Eq.~\eqref{eq:ham_rydberg}.
We consider both nearest-neighbor~(blue) and full van der Waals interactions~(red) with parameters $V_0=12\, \Omega$ and interchain spacing $\tilde{a}\sim 1.51$.
Remarkably, for nearest-neighbor interactions, the oscillations are robust, persisting well beyond the local thermalization timescale $1/\Omega$.
In the limit $V_{0}\to\infty$, the coherent dynamics become exactly periodic with a period $\tau=\pi/V_{0}$ as a consequence of the rainbow tower.
Including long-range interactions leads to faster relaxation dominated by next-nearest-neighbor terms on a timescale $1/V_{i,i+2}$.
This dynamical behavior is confirmed by measuring the average expectation value between inversion partners, $\overline{\langle \sigma^x_i(t) \sigma^x_{\tilde i}(t)\rangle}=\sum_{i}\langle \sigma^x_i(t) \sigma^x_{\tilde i}(t)\rangle/2N$.
Interestingly, the sub-volume-law scars of the PXP model~\cite{scartheo2, scartheo2.1} coexist with the rainbow scars, still displaying a strong dynamical signature, illustrated in Fig.~\ref{fig:panel3}(d) by preparing the system in the N\'eel state.
We emphasize that the dynamical signature of the rainbow tower is more robust than that of the PXP scars for nearest-neighbor interactions.
This results from the fact that $\ket{Z}$ has unit overlap with the rainbow tower in the limit $V_{0}\to\infty$, whereas the PXP tower remains approximate in this limit.
In~\cite{smaterial} we explore various perturbations to Eq.~\eqref{eq:ham_rydberg}, as well as a translation-invariant model in which a similar dynamical signature is found.

\textit{Experimental preparation.}---Rainbow state preparation requires non-local gates to entangle inversion partners at sites $i$ and $\tilde{i}$, posing an experimental challenge.
Recently, however, the rainbow state was prepared in trapped ion quantum simulator~\cite{exp_tfd}.
We recognize that these systems are able to apply nonlocal two-body entangling gates, allowing for easier preparation, but experimental groups are attempting to implement similar gates in Rydberg arrays.
A possible solution is quantum state reversal~\cite{statetrans1,statetrans2,statetrans3}.
Alternatively, in a ladder geometry, the Rydberg system becomes translation-invariant, and state preparation is local.
In~\cite{smaterial} we find the non-ergodic dynamics to persist in this geometry.

\textit{Conclusion.}---This work gives a general recipe to realize a new class of QMBS, dubbed rainbow scars, that are related to the infinite-temperature thermofield double states.
Rainbow scars emerge in any system of the form \eqref{eq:general_construction}, provided $(i)$  $H_{2}=-\mathcal{M}H^{*}_{1}\mathcal{M}$
and $(ii)$ Eq.~\eqref{eq:rainbow_state} is an eigenstate of the coupling $V_{\text{c}}$.
Symmetries enrich the construction, leading to multiple or even towers of rainbow scars with a rich group structure.
These non-thermal states display volume-law entanglement for random bipartitions and sub-volume law scaling for a fine-tuned bipartition, as well as perfect coherent dynamics in the presence of towers.
Our work serves as an experimental blueprint for Rydberg simulators, where we find a robust dynamical signature distinct from previous studies.

\textit{Acknowledgments.}---We thank Adam Kaufman for insightful discussions on the experimental possibilities in the Rydberg system.
Z.-C.Y.~and A.V.G.~acknowledge funding by AFOSR, AFOSR MURI, NSF PFCQC program, DoE ASCR Quantum Testbed Pathfinder program (award No.~DE-SC0019040), U.S.~Department of Energy Award No. DE-SC0019449, DoE ASCR Accelerated Research in Quantum Computing program (award No.~DE-SC0020312), ARO MURI, and DARPA SAVaNT ADVENT.
T.I.~acknowledges the hospitality of the Aspen Center for Physics, which is supported by National Science Foundation grant PHY-1607611.
Portions of this research were conducted with the advanced computing resources provided by Texas A\&M High Performance Research Computing.

\bibliography{references}
\end{document}


\title{Supplementary Material for ``Rainbow Scars: From Area to Volume Law"}

\author{Christopher M. Langlett}
\affiliation{Department of Physics \& Astronomy, Texas A\&M University, College Station, Texas 77843, USA}

\author{Zhi-Cheng Yang}
\affiliation{Joint Center for Quantum Information and Computer Science, NIST/University of Maryland, College Park, Maryland 20742, USA}
\affiliation{Joint Quantum Institute, NIST/University of Maryland, College Park, Maryland 20742, USA}

\author{Julia Wildeboer}
\affiliation{Department of Physics \& Astronomy, Iowa State University, Ames, Iowa 50011, USA}

\author{Alexey V. Gorshkov}
\affiliation{Joint Center for Quantum Information and Computer Science, NIST/University of Maryland, College Park, Maryland 20742, USA}
\affiliation{Joint Quantum Institute, NIST/University of Maryland, College Park, Maryland 20742, USA}

\author{Thomas Iadecola}
\email{iadecola@iastate.edu}
\affiliation{Department of Physics \& Astronomy, Iowa State University, Ames, Iowa 50011, USA}

\author{Shenglong Xu}
\email{slxu@tamu.edu}
\affiliation{Department of Physics \& Astronomy, Texas A\&M University, College Station, Texas 77843, USA}

\begin{abstract}
\end{abstract}
\maketitle
\section{Rainbow scar entanglement entropy}
\label{sec:Entanglement}
In this Appendix, we show that, for a random partition of the system into sub-regions $A$ and $B$, the average entanglement between $A$ and $B$ for the rainbow state $\ket{I}$ scales extensively with the size of the smaller subregion.
(Without loss of generality, we assume region $A$ to be the smaller of the two sub-regions.)
We further study the scaling of the R\'enyi entropy for the projected rainbow scar states of the U(1) tower in the limit of large system size $N$.
We give results for both the standard entanglement cut and a fine-tuned cut for which the rainbow state has zero entanglement.
We emphasize that the results of this Appendix also hold for the other rainbow states $\ket{X},\ket{Y},$ and $\ket{Z}$, since these states are obtained from $\ket{I}$ by unitary operations that generate no additional entanglement.

\subsection{Average Entanglement Entropy for a Random Bipartition}
We consider the rainbow state $\ket{I}$~[see Eq.~(2) in the main text] in a system of $2N$ sites.
In total there are $2^{2N}$ possible bipartitions, since each site can be either included or excluded from region $A$.
The size of region $A$ for a given bipartition is $\ell=2\, n_{\text{bp}}+n_{\text{s}}$ where $n_{\text{bp}}$ is the number of Bell pairs enclosed in region $A$ and $n_{\text{s}}$ is the number of singleton sites (or, equivalently, the number of entanglement ``bonds" cut by the bipartition).
Given a bipartition, the entanglement entropy scales with the number of singletons, $S=n_{\text{s}}\log(d)$~(for concreteness we set $d=2$.).
For each $\ell \in [0, 2N]$, we determine the average singleton number, $\overline{n}_{\text{s}}$, as follows:
\bea
\label{eq:av_singleton}
\overline{n}_{\text{s}} & =\sideset{}{'}\sum^{\ell}_{n_{\text{s}}=0} n_{\text{s}} P_{\ell}(n_{\text{s}})
\eea
where $P_{\ell}(n_{\text{s}})$, the probability distribution of $n_{\text{s}}$ for fixed $\ell$, satisfies
\bea
\sideset{}{'}\sum_{n_{\text{s}}=0}^{\ell} P_{\ell}(n_{\text{s}})=1.
\eea
Here, the prime on the summation symbol denotes that the sum runs only over the values of $n_{\rm s}$ for which $n_{\rm bp}=(\ell-n_{\rm s})/2$ is an integer.
$P_{\ell}(n_{\text{s}})$ takes the combinatorial form
\bea
\label{eq:pdist}
P_{\ell}(n_{\text{s}}) =\frac{1}{\binom{2N}{\ell}} \binom{N}{n_\text{bp}}\binom{N-n_\text{bp}}{n_{\text{s}}}2^{n_{\text{s}}}.
\eea
The above expression is determined first by picking $n_{\text{bp}}$ from the total number of $N$ Bell pairs in the rainbow state.
The remaining $N-n_{\rm bp}$ Bell pairs furnish the $n_{\rm s}$ singletons.
The factor $2^{n_{\rm s}}$ arises from the fact that each singleton site can reside within either of subsystems 1 and 2.
The remaining factor of $\binom{2N}{\ell}$ ensures normalization.
Combining Eq.~\eqref{eq:av_singleton} and Eq.~\eqref{eq:pdist} results in the bipartition-averaged entanglement entropy
\bea
S_{\text{av}} = \frac{1}{2N-1}(2N-\ell)\ell \log(2),
\eea
which fits the numerical result in Fig.~1(a) of the main text.
Note that the above expression for $S_{\rm av}$ scales extensively with system size $N$ when $\ell\propto N$. 
In the large-$N$ limit the probability distribution $P_{\ell}(n_{\text{s}})$ approaches a Gaussian distribution of the form,
\bea
P_{\ell}(n_{\text{s}}) \rightarrow \sqrt{\frac{2N}{\bar n_{\text{s}}^2 \pi} }\exp \left( -N\frac{(n_{\text{s}}-\bar n_{\text{s}})^2}{2\bar n_s^2}\right),
\eea
where the mean
\bea
\bar n_{\text{s}} =\frac{1}{2N} (2N-\ell)\ell.
\eea
The standard deviation of $n_{\text{s}}/N \rightarrow 0$ as $N\to\infty$, indicates that the ratio $n_{\text{s}}/N$ takes the average value for a typical bipartition. 
This result emphasizes that the entanglement scaling of the state $\ket{I}$ for a typical entanglement cut is extensive, in stark contrast with previous exact constructions of scar states.
\begin{figure}
    \includegraphics[width=\columnwidth]{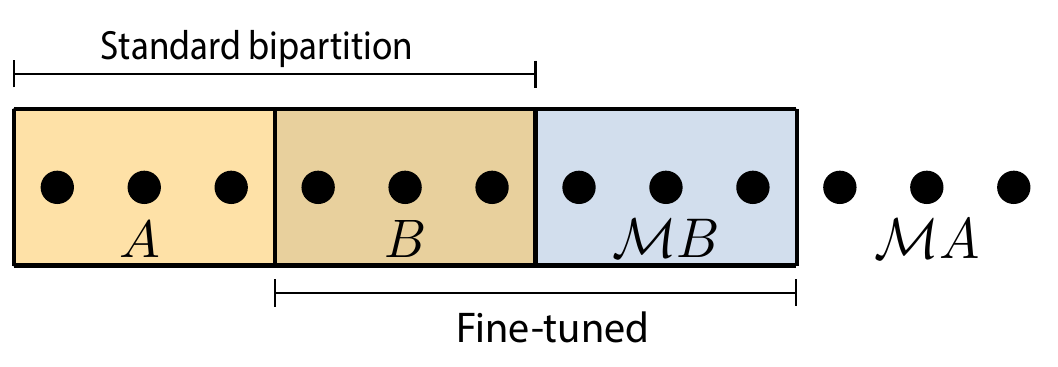}
    \caption{\textit{Entanglement Bipartitions.}
    The standard bipartition~(orange) is constructed with a cut placed between sites $N$ and $(N+1)$ and spanned by $\ket{S^{A(B)}}$ in the local $S^{z}$ basis.
    The fine-tuned bipartition~(blue) is formed by a cut between sites $\left(N/2,N/2+1\right)$, as well as, $\left(3N/2,3N/2+1\right)$ and spanned by the basis states $\ket{S^{B}}$ and its mirror $\ket{\I S^{B}}$.}
    \label{fig:panel1_A}
\end{figure}

\subsection{Entanglement of the U(1) Rainbow Tower}
In this section we consider the rainbow scars of the XYZ model with a U(1) symmetry [Eq.~(3) in the main text with $h_x=h_y=0$ and $J_{x}=J_{y}$] and perform a large-$N$ analysis of the R\'enyi entropy.
When the total magnetization $S^{z}=\sum^{2N}_{i=1} S^z_i$ of the combined system is conserved, the Hilbert space is a direct sum of sub-sectors labelled by $S^{z}$ eigenvalues $\{-N,\cdots,N\}$.
We represent the rainbow scars within each magnetization sector as:
\bea
\label{eq:charge_rainbow}
\ket{\Phi_{n}} & = \mathcal{N}(n) \left( J^{+}\right)^{n} \prod_{i=1}^{2N}\ket{\downarrow}_{i},\\
& = \binom{N}{n}^{-1/2} \sum_{S} \ket{S}\otimes \ket{\mathcal{M} S},
\eea
where $\ket{S}$ is in the local $S^{z}$ basis for the half-chain with total magnetization $m_{n}=n-N/2$ with $n \in [0,\cdots,N]$.
Importantly, the state $\ket{\Phi_{n}}$ is the sum over all permutations of $n$ mirror excitations in a polarized background~(i.e., raised spins at sites $i$ and $\tilde{i}$).
We emphasize that with each application of $J^{+}$ the number of excitations~(raised spins) increases by two, resulting in the rainbow state having finite projection onto \textit{every other} magnetization sector, leaving a tower of $(N+1)$ states.
By contrast, previously studied U(1) scar towers have a non-thermal eigenstate within each magnetization sector~\cite{scarsym3,scartheo5}.

\subsubsection{Standard Cut}
We first consider the ``standard" bipartition, where the entanglement cut is placed between sites $N$ and $N+1$.
The state~\eqref{eq:charge_rainbow} is already in Schmidt-decomposed form with Schmidt coefficients
\bea
\lambda = \binom{N}{n}^{-1/2},
\eea
each with multiplicity $\binom{N}{n}$, ensuring the Schmidt coefficients are properly normalized.
Therefore, the entanglement entropy takes the following form:
\bea
\label{eq:S-standard}
S = \log \binom{N}{n}\rightarrow -N\left((1-\gamma)\log(1-\gamma)+\gamma \log \gamma\right),
\eea
where $\gamma = n/N$.
Thus, for the standard cut the entanglement entropy scales extensively with system size, in contrast with previous examples of exact U(1) scar towers.
Indeed, Eq.~\eqref{eq:S-standard} is the maximum possible entanglement between two quantum systems with Hilbert space dimension $\binom{N}{n}$.

\subsubsection{Fine-Tuned Cut}
We consider the state $\ket{\Phi_{n}}$ in a system of $2N$ sites, which we bipartition into regions $A$ and $B$ with sizes $N_{A}$ and $N_{B}=2N-N_{A}$.
Here we focus on bipartitions of equal size, i.e., $N_{A}=N_{B}=N$~(we take $N$ to be even.).
Specifically, we focus on the fine-tuned bipartition where cuts are placed between sites $N/2$ and $(N/2+1)$, and between sites $3N/2$ and $(3N/2+1)$, which identifies the middle half of the system as region $A$.
The entanglement spectrum is completely characterized by the Schmidt coefficients, which are found by first decomposing the state~\eqref{eq:charge_rainbow} as
\bea
\ket{\Phi_{n}} = \sum_{k}\lambda_{k}\ket{\Phi_{k}^{A}}\ket{\Phi_{n-k}^{B}},
\eea
where $\ket{\Phi_{j}^{A(B)}}$ are a set of orthonormal states for region $A(B)$ in the local $S^{z}$ basis, labelled by the number $j$ of mirror excitations, given by
\bea
\ket{\Phi_k^A}=\binom{N/2}{k}^{-1/2}\sum_{S^A} \ket{S^A} \ket{\I S^A} \\ \ket{\Phi_{n-k}^B}=\binom{N/2}{n-k}^{-1/2}\sum_{S^B}\ket{S^B} \ket{\I S^B}.
\eea
The sum in $\ket{\Phi_k^A}$ is over all states $\ket{S^A}$ with magnetization $m_{k}=k-N/4$ in region $A$ and the sum in $\ket{\Phi_k^B}$ is over all states $\ket{S^B}$ with magnetization $m_{n-k}=(n-k)-N/4$ in region $B$.
The Schmidt coefficients $\lambda_{k}$, properly normalized, are given by
\bea
\label{eq:sch_coeff}
\lambda^{2}_{k} = \frac{\binom{\frac{N}{2}}{k}\binom{\frac{N}{2}}{n-k}}{\binom{N}{n}},
\eea
and satisfy $\sum_{k=0}^{n}\lambda_{k}^{2}=1$.
Determining the $\lambda_{k}$ permits the construction of the R\'enyi entropy of order $\alpha$ defined as:
\bea
S^{(\alpha)}=\frac{1}{1-\alpha} \log\left(\sum_{i}\lambda_{i}^{2\alpha}\right),
\eea
The R\'enyi entropy is then computed by taking the logarithm of the following result,
\bea
\label{eq:sch_coeff}
e^{(1-\alpha)S^{(\alpha)}} =\sum_{k=0}^{n} \frac{\binom{\frac{N}{2}}{k}^{\alpha}\binom{\frac{N}{2}}{n-k}^{\alpha}}{\binom{N}{n}^{\alpha}}.
\eea
Using saddle point methods, the second-order~($\alpha=2$) R\'enyi entropy in the large-$N$ limit has the scaling form
\bea
\label{eq:entropy}
S^{(2)} \underset{N\rightarrow \infty}{=} \frac{1}{2}\log\left(N\pi\gamma(1-\gamma)\right),
\eea
We note that this result is different than in the case of other symmetries such as $\mathbbm{Z}_{2}$, where the fine-tuned cut has zero entanglement; here, the scar state in each magnetization sector scales logarithmically with $N$ provided $\gamma=n/N$ is finite.
\section{Heisenberg Perturbation}
\label{sec:HeisPert}
We consider the consequence of perturbing the U(1)-symmetric point of two coupled Heisenberg XYZ chains composed of $N$ spins.
In the main text~[see Fig.~3(d)] we demonstrated a perturbation of the form $V_{\text{pert}}=\mathcal{O}\otimes\mathbbm{1} - \mathbbm{1}\otimes\mathcal{O}^{*}$, which respects the structure of Eq.~(1) but breaks the U(1) symmetry.
In this instance the oscillations remain robust for perturbations strengths up to $D\sim 0.5$ before thermalization rapidly sets in.
\begin{figure}[h]
    \centering
    \includegraphics[width=0.75\columnwidth]{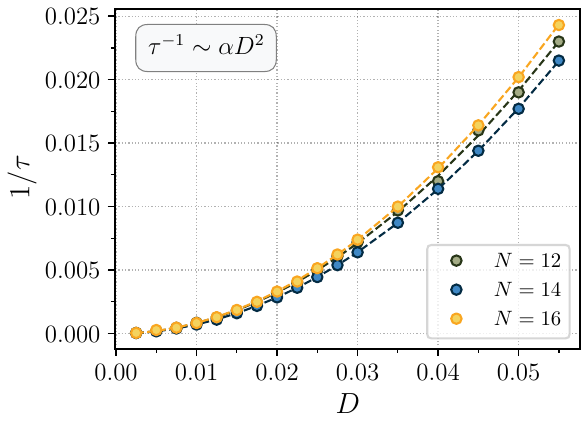}
    \caption{
    (a) \textit{Perturbation of Heisenberg XYZ model.}
    The inverse lifetime is extracted from fitting $\langle (S^{x}(t))^{2}\rangle/N$ with the function $f(\tau)=Ae^{-t/ \tau}\cos(2\mu t)+1/2$.
    The constant $\alpha \sim 6.82$ is found to be nearly twenty times larger than for the perturbation considered in the main text.
    The parameters used are $J_{z}=2.0, J_{x}=J_{y}=1.0,  \mu=0.5, \Tilde{J}=0.5,\lambda_{\text{c}}=1.5,dt=0.1$.
    }
    \label{fig:panel2_C}
\end{figure}
We emphasize that in previous studies of perturbations of scarred models~\cite{scarpert1} the oscillations decay at much lower perturbation strengths.
Therefore, the structure alone of Eq.~(1) seems to act as a stabilizer of the scars' coherent dynamics without needing to add extra terms to the Hamiltonian.
We confirm this suspicion by considering a perturbation which explicitly breaks the structure of Eq.~(1) and the U(1) symmetry.
Specifically, consider a perturbation of the form $V_{\text{pert}}=\mathcal{O}\otimes\mathbbm{1} + \mathbbm{1}\otimes\mathcal{O}^{*}$ with the operator $\mathcal{O}=\sum_{i}^{N}S_{i}^{x}S_{i+1}^{x}$.
We extract the inverse lifetime of the oscillations' decay by fitting to a function of the form $f(\tau)=Ae^{-t/\tau}\cos(\mu t)+1/2$ and plot it against the perturbation strength $D$ in Fig.~\ref{fig:panel2_C}.
The inverse lifetime is then fit with the function $\tau^{-1}=\alpha D^{2}$, which is expected from Fermi's golden rule.
Surprisingly, while the perturbation considered here follows the same power law as in the main text, here the coefficient $\alpha=6.82$ is roughly twenty times larger than the value ($\alpha=0.4$) obtained for the perturbation considered in the main text.
Due to the size of this constant, thermalization sets in at perturbation strengths ten times smaller than when the construction's structure is respected.
In the future, we plan on exploring the role the structure plays in dynamics and obtain a better understanding of the constant $\alpha$.

\section{Rydberg System}
In this appendix we first study the Rydberg system from the main text in the presence of different experimental errors.
For example, in order to ensure that the central sites are at the optimal detuning value, very precise laser placement is required, which generically is not perfectly accurate.
As a result, some of the applied light will be imparted on the atoms directly adjacent to the two central ones, giving them a finite detuning.
We study this scenario in depth and find the rainbow scar state to be robust over a wide range of parameters.

We further address the problem of state preparation by considering a the Rydberg system in the geometry of a ladder rather than a chain.
Here the rainbow state is prepared on each rung of the ladder, which can be achieved with local entangling gates.
Another reason why the ladder geometry is appealing is that its Hamiltonian is translation invariant.

\subsection{Experimental Error}
We now study the effects of various possible sources of error that may occur in an experimental realization of the model discussed in the main text.
\label{sec:RydPert}
\begin{figure}[h]
    \centering
    \includegraphics[width=\columnwidth]{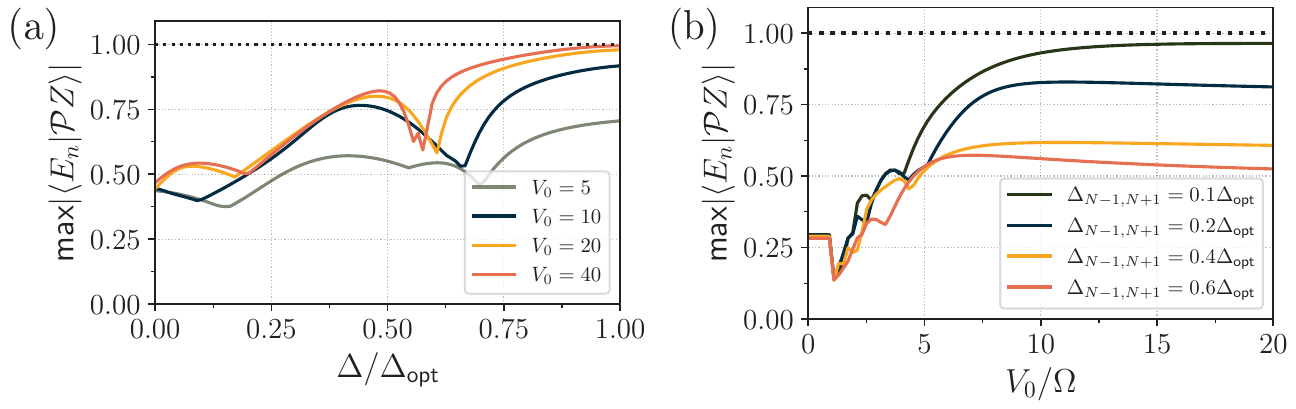}
    \caption{\textit{Rydberg Perturbation.}
    (a) Maximum overlap of $\ket{\mathcal P Z}$, the projection of $\ket{Z}$ into the sub-sector absent of neighboring Rydberg excitations, with all eigenstates in that sub-sector as a function of the detuning $\Delta$ on the two central sites.
    As $\Delta$ approaches $\Delta_{\rm opt}$ and the nearest-neighbor interaction strength $V_0$ is increased, the maximum overlap approaches unity.
    (b) Same as (a) with the detuning on the central sites fixed to $\Delta_{\text{opt}}$, while the detuning on the sites directly adjacent to the central sites is fixed to different fractions of the optimal value. In this case, the maximal overlap is plotted as a function of the nearest-neighbor interaction strength $V_0$.
    Parameters used in (a), (b): $\Omega / 2\pi = 2$MHz, $\Delta_{\text{opt}} = V_{0}/2 \tilde{a}^{6}$ with $\tilde{a}\sim 1.51$ and $2N=12$.}
    \label{fig:panel2_B}
\end{figure}

As an experimental realization of rainbow scars, we studied two coupled chains of Rydberg atoms.
To obtain the strongest dynamical signature of the rainbow scars, each atom should be on resonance, i.e, $\Delta_{i}=0$, except for the atoms on the two central sites, where the detuning takes an optimal value $\Delta_{\rm opt}$.
In practice, satisfying these conditions is challenging; therefore, in this appendix we study how robust the scar state is under more realistic conditions.

The first possibility we consider is to have the two central sites not set exactly at the optimal detuning; we can then define their detuning to be $\Delta = \eta \Delta_{\text{opt}}$, where $\eta \in [0,1]$.
The case $\eta=0$ corresponds to the entire chain being on resonance, making the chain translation-invariant except for the spacing between the two central sites.
To probe the role of the detuning on the central sites, in Fig.~\ref{fig:panel2_B}(a) we plot the maximum overlap of $\ket{\mathcal P Z}$, the projection of the rainbow state $\ket{Z}$ into the sub-sector absent of neighboring Rydberg excitations, with all eigenstates in that sub-sector as a function of the detuning $\Delta$ on the two central sites.
Surprisingly, even in the case $\eta=0$ the maximum overlap is around fifty percent~[see Fig.~\ref{fig:panel2_B}(a)], demonstrating that the signature of the scar state is \textit{not} strictly conditioned on the detuning on the central sites being optimal.
As the ratio $\Delta/\Delta_{\rm opt}$ approaches one and the interaction strength $V_0$ approaches infinity, the maximum overlap converges to unity in agreement with our findings in the main text.

Another possibility is error due to the coupling laser not being entirely focused on the central two sites, thereby imparting a non-zero detuning onto the sites directly adjacent to them.
In Fig.~\ref{fig:panel2_B}(b) we re-calculate the maximal overlap between $\ket{\mathcal P Z}$ and the exact eigenstates with the detuning on sites $N-1$ and $N+2$ set to various fractions of $\Delta_{\rm opt}$.
We find that, even at moderately large fractions of $\Delta_{\rm opt}$, the maximum overlap remains above fifty percent provided the nearest-neighbor interaction is sufficiently large.
From an experimental perspective, these results suggest that the dynamical signature of the rainbow scars will emerge for a wide range of detunings.

\subsection{Rydberg Ladder}
\label{sec:Ladder}
\begin{figure}[h]
    \centering
    \includegraphics[width=\columnwidth]{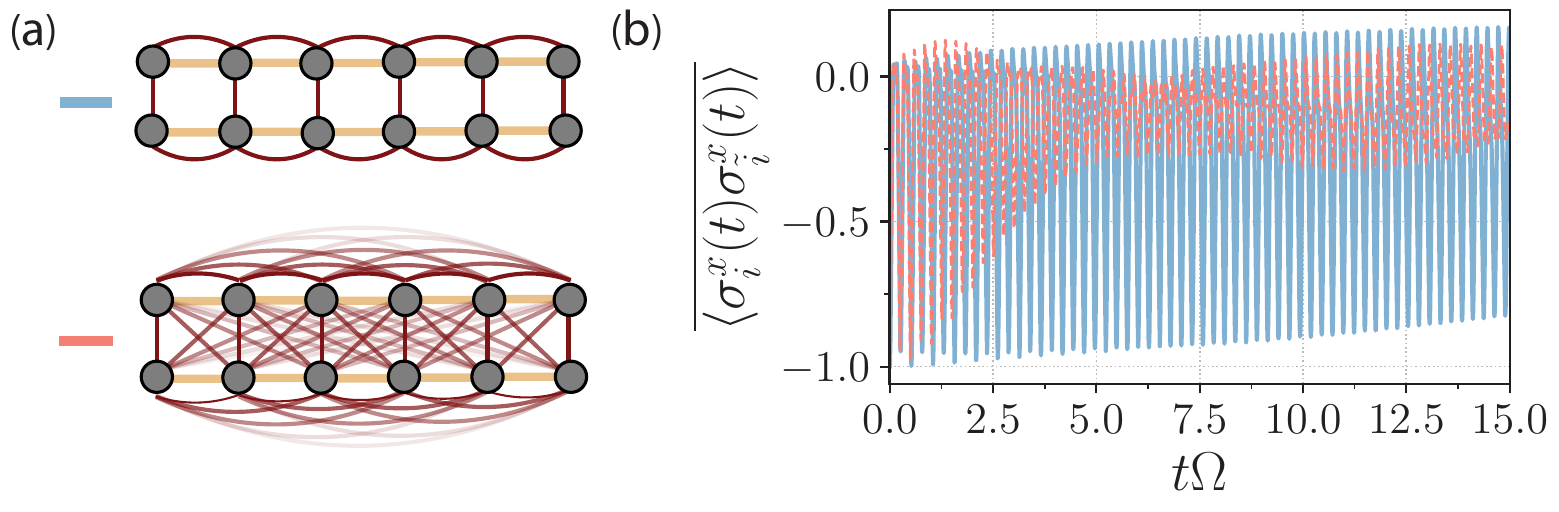}
    \caption{\textit{Rydberg Ladder.}
    (a) Cartoon depiction of the Rydberg ladder considered in Sec.~\ref{sec:Ladder} for nearest-neighbor and all-to-all interactions.
    (b) Dynamics of the correlator $\overline{\langle \sum \sigma^x_{i,1} \sigma^x_{i,2}\rangle}$ for the initial state $\ket{Z}$, calculated using Krylov time-evolution.
    The parameters are: $2N=16$, $\Omega / 2\pi = 2$MHz, $V_{0}=12\Omega$, $\Delta_{\text{opt}}=V_{0}/2\tilde{a}^{6}$, and $\tilde{a}\sim 1.51$.}
    \label{fig:panel2_A}
\end{figure}
In the main text we showed that, when a non-uniformly spaced Rydberg chain has its two central atoms detuned to a specific value, they become coupled by an Ising interaction, resulting in non-ergodic dynamics from a tower of rainbow scar states.
While the rainbow state has a strong dynamical signature, its experimental preparation can be difficult.
A possible resolution mentioned in the main text is to ``fold" the chain into a ladder, which permits the use of local gates for state preparation, as well as rendering the model translation invariant.
Below, we give numerical evidence that the non-ergodic signature of the rainbow scars persists in the ladder geometry under experimentally reasonable conditions.
To this end, we begin with the Hamiltonian 
\bea
H_{\parallel} & = \sum^2_{b=1}\left(\frac{\Omega}{2}\sum_{i=1}^{N}\sigma^{x}_{i,b} + \sum_{i<j}V_{i,j}n_{i,b}n_{j,b}\right),\\
H_{\perp} & = -\sum^N_{i=1}\sum^2_{b=1}\Delta_{i,b}n_{i,b} + \sum_{i,j} \tilde{V}_{i,j}n_{i,1}n_{j,2},
\label{eq:rydberg_ladder}
\eea
where $b=1,2$ labels the legs of the ladder.
We set the interatomic spacing $a=1$ between atoms on the same leg, and define $\tilde{a}$ to be the spacing between the legs.
The operator $\sigma^{x}_{i}$ connects the internal ground state $\ket{g}_{i}$ to the Rydberg state $\ket{r}_{i}$ of the $i$-th atom, with parameters $\Omega$~(Rabi frequency) and $\Delta_i$~(detuning) characterizing the drive laser.
Rydberg atoms in the same leg interact through $V_{i,j}=V_{0}/r_{i,j}^{6}$, with operators $n_{i}=\left(1+\sigma^{z}_{i}\right)/2$.
Rydberg atoms in different legs interact through $\tilde V_{i,j}=V_{0}/\tilde r_{i,j}^{6}$, where $\tilde r_{i,j}$ is the distance between site $i$ in the leg $b=1$ and site $j$ in the leg $b=2$.
In the limit $V_{i,i+1}\gg\Omega\gg V_{i,i+2}$, we take $\tilde V_{i,i}=V_{0}/ \tilde{a}^{6}$ to be comparable to $\Omega$; equivalently, we take $\tilde{a} > 1.0$.
By contrast to the non-uniformly spaced 1D chain, where only the middle sites are off resonance, here each rung pair is detuned to the optimal value, $\Delta_{i,1}=\Delta_{i,2}=\Delta_{\text{opt}}=\tilde V_{i,i}/2$.
With this detuning, each rung pair interacts through an Ising coupling, $V_0\sigma_{i,1}^z\sigma_{i,2}^z/4\tilde{a}^{6}$.
In the strong-coupling limit $V_{i,i+1}\gg \Omega \gg V_{i,i+2}$, the Hilbert space splits into the sub-sectors discussed in the main text.

In this ladder geometry, the equally spaced tower of states discussed in the main text still reveals itself through the system's dynamics.
We probe the presence of the tower by preparing the ladder in the $\ket{Z}$ rainbow state and, using experimentally reasonable parameters, simulate the dynamics well beyond the local relaxation timescale, $1/\Omega$.
In~Fig.~\ref{fig:panel2_A}(b), we measure the expectation value $\overline{\langle \sum \sigma^x_{i,1} \sigma^x_{i,2}\rangle}$ for the case of both nearest-neighbor and all-to-all long-range interactions with parameters $V_0=12\Omega$ and $\tilde{a}\sim1.51$.

In the coupled-1D-chain example discussed in the main text, the two chains interact through a single term on the center sites.
Here, instead, there are $N$ Ising couplings between the legs for nearest-neighbor interactions.
Remarkably, the non-ergodic dynamics remain robust to this increase in interactions, which results from the projection of the $\ket{Z}$ rainbow state onto each sub-sector being a local eigenstate of $H_\perp$ in the strong coupling limit.
Introducing long-range interactions leads to faster decay, except here the primary perturbation comes from the diagonal interaction between legs, rather than next-nearest-neighbor interactions within each leg.
Despite the fact that the two sub-systems are coupled by more than a single term, the non-ergodic dynamics persists.
The ladder geometry thus provides a promising alternative way to probe experimentally the dynamical signature of rainbow scars.

\bibliography{references}